\documentclass[12pt]{article}

\usepackage{graphicx}
\usepackage{amsfonts,amsmath,amssymb,amsthm}
\usepackage{indentfirst}
\usepackage{cite,hyperref}
\usepackage{mathptmx}

\numberwithin{equation}{section}
\DeclareMathAlphabet{\mathcal}{OMS}{cmsy}{b}{n}

\begin{document}

\title{On the Bhabha scattering for $z=2$ Lifshitz QED}
\author{R. Bufalo$^{1,2}$\\
\textit{{$^{1}${\small Department of Physics, University of Helsinki, P.O. Box 64}}}\\
\textit{\small FI-00014 Helsinki, Finland}\\
\textit{{$^{2}${\small Instituto de F\'{\i}sica Te\'orica (IFT), Universidade Estadual Paulista}}} \\
\textit{\small Rua Dr. Bento Teobaldo Ferraz 271, Bloco II Barra Funda, CEP
01140-070 S\~ao Paulo, SP, Brazil}\\
}
\maketitle
\date{}

\begin{abstract}
In this paper we compute and discuss the differential cross section of the Bhabha scattering in the framework of the $z=2$ Lifshitz QED. We start by constructing the classical solutions for the fermionic fields, in particular the completeness relations, and also deriving the theory's propagators. Afterwards, we compute the photon exchange and pair annihilation contributions for the Bhabha's process, and upon the results we establish the magnitude of the theory's free parameter by looking for small deviations of the QED tree results.
\end{abstract}

\newpage

\section{Introduction}

Within the field theory context the Lifshitz models, based on an anisotropy between space and time \cite{ref16}, i.e. the scaling $x^i\rightarrow \Lambda x^i$  whereas $t\rightarrow \Lambda^z t$, always had a variety of applications permeating the literature of condensate matter physics, and nowadays has been reviver and gained much attention within the context of the search for a perturbatively consistent theory for gravity \cite{ref17}. The remarkable advantage of this approach is mainly motivated by the possibility of defining new renormalizable interactions, but they are constructed in such a ways to avoids the appearance of \emph{ghosts} (negative energy modes) whose presence is an imprint of theories with higher time derivatives.

Although these Lifshitz field theories have a better ultraviolet behavior at the expenses of breaking Lorentz
invariance, many interesting and rich features regarding the theory's renormalizability \cite{ref1} and also in discussions on Lorentz symmetry violation \cite{ref2} have been observed. In exploring implications of the time-space asymmetry of a given field theory's renormalizability, many investigations were undertaken and reported recently, for instance, studies considering Lifshitz gauge field theories \cite{ref18,ref19a,ref19b}, in particular a detailed study of its general properties can be found in the Ref.\cite{ref4a,ref4b}; moreover, scalar field theories \cite{ref4a,ref4b,ref23}, four-fermion theory \cite{ref3a,ref3b}, and $CP^{N - 1}$ model \cite{ref24} were also considered. In completion, we may also cite the study of formal aspects of Lifshitz field theories according to the BPHZ scheme \cite{ref25}.

Recently, an interesting study of a phenomenological viable $z=2$ Lifshitz modified quantum electrodynamics (QED) has been reported \cite{ref6}. By the fact that the theory is super-renormalizable, non-trivial aspects of its quantum content were discussed. Nevertheless, the possible effects on scattering processes in the framework of QED by these new super-renormalizable couplings have not been investigated. Some attention has been paid in studying scattering process, in particular to the Bhabha scattering, in the Lorentz violating field theories, due to the emergence of new vertices in the theory \cite{ref14,ref7,ref8}, and also at the generalized electrodynamics \cite{ref9}. In these studies, experimental data was used in order to obtain a bound value for the theories free parameters.

Remarkably, Bhabha scattering is one of the most fundamental reactions in QED processes \cite{ref11} and has played an important role in experiments in colliders \cite{ref26a,ref26b,ref26c}, since it is particularly used to determine the luminosity
of the collisions \cite{ref10}. It is important to emphasize that nowadays there are ongoing theoretical analysis present in the literature \cite{ref12a,ref12b,ref12c}, in the search of an improvement in the calculations of higher-order contributions for the Bhabha's cross section. Therefore, motived by theoretical and experimental grounds, we believe to be interesting and rather natural to consider and analyze the behavior of the differential cross section for Bhabha scattering in the framework of the $z=2$ Lifshitz QED.

In this paper, we discuss a scattering process in the framework of $z=2$ Lifshitz modified quantum electrodynamics. In Sec. \ref{sec:1} we start by making a brief review of the $z=2$ QED and then we construct in details the classical solutions for the fermionic fields, as well as the completeness relations for the spinors, and also determine the theory's fermionic and gauge field propagators quantities that are all needed in the calculation of the cross-section for the Bhabha scattering.
Next, in Sec. \ref{sec:2}, we discuss and present a detailed calculation of the S-matrix elements, the transition amplitude for the theory. The calculation is performed by taking into account the high energy limit, i.e. we take the fermionic mass to be zero; moreover, we express the kinematic variables in the center-of-mass frame. In conclusion, we establish the magnitude of the theory's free parameter by looking for small deviations of the QED tree results \cite{ref10}. In Sec. \ref{sec:3} we summarize the results, and present our final remarks and prospects.

\section{General discussion}
\label{sec:1}

In this section we will introduce our basic notation and describe the model and its classical solutions. Let us consider here the following Lagrangian density for the $z=2$ Lifshitz QED \cite{ref6},
\begin{equation}
\mathcal{L}=\frac{1}{2}F_{0i}F_{0i}-\frac{1}{4}F_{ij}\left(M^{2}-\Delta\right)F_{ij}
+\overline{\psi}\left(i\gamma_{0}D_{0}-iM\gamma_{k}D_{k}-D_{k}D_{k}-m^{2}\right)\psi,\label{eq: 0.1}
\end{equation}
with the metric signature $\left(+,-,-,-\right)$, the covariant derivative $D_{\mu}=\partial_{\mu}+igA_{\mu}$, and the field stress tensor $F_{\mu\nu}=\partial_{\mu}A_{\nu}-\partial_{\nu}A_{\mu}$. This theory is known to be super-renormalizable, but still contains power counting diverging graphs, which were shown to be actually UV finite \cite{ref6}. From the Lagrangian density \eqref{eq: 0.1} we may read that the length dimensions are, in $3+1$ dimensions,
\begin{align}
\left[A_{0}\right] & =\left[\psi\right]=L^{-\frac{3}{2}},\quad\left[A_{i}\right]=L^{-\frac{1}{2}},\nonumber \\
\left[g\right] & =L^{-\frac{1}{2}},\quad\left[m\right]=\left[M\right]=L^{-1}.
\end{align}
The theory is invariant under the $U\left(1\right)$ gauge symmetry
\begin{equation}
\psi\rightarrow e^{i\sigma}\psi,\quad A_{\mu}\rightarrow A_{\mu}+\frac{1}{g}\partial_{\mu}\sigma.
\end{equation}
We are interested here in evaluate the Bhabha's differential cross-section, and for that matter we need to construct the solution for the fermionic fields \cite{ref13}. Hence, in order to construct these solutions for the Dirac's fields we shall consider the following  modified free field equation,
\begin{equation}
\left(i\gamma_{0}\partial_{0}-iM\gamma_{k}\partial_{k}-\partial_{k}\partial_{k}-m^{2}\right)\psi=0,\label{eq: 0.2}
\end{equation}
the coefficients of the differential equation are constants, thus $\psi\left(x\right)=e^{-ipx}\chi\left(p\right),
$ will be a solution. Hence, we obtain
\begin{equation}
\left(\gamma_{0}p_{0}-M\gamma_{k}p_{k}+p^{2}-m^{2}\right)\chi\left(p\right)=0,\label{eq: 0.3}
\end{equation}
where $p^{2}=p_{i}p_{i}$. For the energy eigenvalues, we have
\begin{equation}
p_{0}\equiv\pm E_{p}=\pm\sqrt{M^{2}p^{2}+\left(p^{2}-m^{2}\right)^{2}},\label{eq: 0.4}
\end{equation}
for each value of $p_{0}$, the solution \eqref{eq: 0.3} has a two-dimensional solution space. For the explicit calculation, we choose
\begin{equation}
\chi\left(p\right)=\left(\begin{array}{c}
u_{s}\left(p\right)\\
v_{s}\left(p\right)
\end{array}\right),
\end{equation}
and using the following representation for the $\gamma$-matrices,
\[
\gamma^{0}=\left(\begin{array}{cc}
\mathbf{1} & 0\\
0 & -\mathbf{1}
\end{array}\right),\quad\vec{\gamma}=\left(\begin{array}{cc}
0 & \vec{\sigma}\\
-\vec{\sigma} & 0
\end{array}\right),
\]
where $\mathbf{1}$ is a two-dimensional identity matrix, and $\vec{\sigma}$ are the set of Pauli matrices. These all lead to,
\begin{align}
\left(p_{0}+p^{2}-m^{2}\right)u_{s}\left(p\right) & =M\left(\vec{\sigma}.\vec{p}\right)v_{s}\left(p\right),\\
\left(p_{0}-p^{2}+m^{2}\right)v_{s}\left(p\right) & =M\left(\vec{\sigma}.\vec{p}\right)u_{s}\left(p\right).
\end{align}
Nonetheless, it is not difficult to show that $u\left(p\right)$ corresponds to those solutions with positive energy, $p_{0}=+E_{p}$, while $v\left(p\right)$ are solutions with negative energy, $p_{0}=-E_{p}$. The next step involves in defining the energy projection operators, in such a way
\begin{align}
\Lambda^{+}\left(p\right)u_{s}\left(p\right) & =u_{s}\left(p\right),\\
\Lambda^{-}\left(p\right)v_{s}\left(p\right) & =v_{s}\left(p\right),
\end{align}
and
\begin{equation}
\Lambda^{-}\left(p\right)u_{s}\left(p\right)=\Lambda^{+}\left(p\right)v_{s}\left(p\right)=0.
\end{equation}
We can then show that the operators
\begin{equation}
\Lambda^{\pm}\left(p\right)=\frac{\mp\gamma_{0}p_{0}\pm M\gamma_{k}p_{k}+p^{2}-m^{2}}{2\left(p^{2}-m^{2}\right)},\label{eq: 0.5}
\end{equation}
satisfy the above relations, as well as
\begin{gather}
\Lambda^{+}\left(p\right)+\Lambda^{-}\left(p\right)  =\mathbf{1},\quad \Lambda^{+}\left(p\right)\Lambda^{-}\left(p\right)  =0,\\
\left[\Lambda^{\pm}\left(p\right)\right]^{2}  =\Lambda^{\pm}\left(p\right).
\end{gather}
Thus, we have the completeness relations satisfied by
\begin{align}
\Lambda_{\alpha\beta}^{+}\left(p\right) & =\sum_{s=1}^{2}u_{\alpha}\left(p,s\right)
\overline{u}_{\beta}\left(p,s\right)=\left[\frac{-\gamma_{0}p_{0}+M\gamma_{k}p_{k}+p^{2}-m^{2}}{2\left(p^{2}-m^{2}\right)}\right]_{\alpha\beta},\label{eq: 0.6a}\\
\Lambda_{\alpha\beta}^{-}\left(p\right) & =-\sum_{s=1}^{2}v_{\alpha}\left(p,s\right)\overline{v}_{\beta}\left(p,s\right)=\left[\frac{\gamma_{0}p_{0}-M\gamma_{k}p_{k}+p^{2}-m^{2}}{2\left(p^{2}-m^{2}\right)}\right]_{\alpha\beta}.\label{eq: 0.6b}
\end{align}
Finally, we have the free solutions are written as
\begin{align}
\psi\left(x\right) & =\sum_{r}\int\frac{d^{3}p}{\left(2\pi\right)^{\frac{3}{2}}}\left(\frac{p^{2}-m^{2}}{E_{p}}\right)^{\frac{1}{2}}\left[b_{r}\left(p\right)u_{r}\left(p\right)e^{-ipx}+d_{r}^{\dagger}\left(p\right)v_{r}\left(p\right)e^{ipx}\right],\label{eq: 0.7a}\\
\overline{\psi}\left(x\right) & =\sum_{r}\int\frac{d^{3}p}{\left(2\pi\right)^{\frac{3}{2}}}\left(\frac{p^{2}-m^{2}}{E_{p}}\right)^{\frac{1}{2}}\left[b_{r}^{\dagger}\left(p\right)\overline{u}_{r}\left(p\right)e^{ipx}+d_{r}\left(p\right)
\overline{v}_{r}\left(p\right)e^{-ipx}\right],\label{eq: 0.7b}
\end{align}
where, the operators algebra read,
\begin{equation}
\left\{b_{r}\left(p\right),b_{s}^{\dagger}\left(q\right)\right\}=\left\{d_{r}\left(p\right)
,d_{s}^{\dagger}\left(q\right)\right\}=\delta_{rs}\delta\left(\vec{p}-\vec{q}\right).\label{eq: 0.8a}
\end{equation}
From such construction one can easily show that the equal-time anti-commutation relations
is satisfied
\begin{equation}
\left\{\psi_{\alpha} (x),\psi_{\beta}^{\dagger} (y)\right\}_{x_0 = y_0}=\delta_{\alpha \beta} \delta^{(3)}\left(\vec{x}-\vec{y}\right).\label{eq: 0.8b}
\end{equation}
Besides, we can also determine the fermion propagator
\begin{equation}
S\left(p_{0},p\right)=i\frac{\gamma_{0}p_{0}-M\gamma_{k}p_{k}+p^{2}-m^{2}}{p_{0}^{2}-M^{2}p^{2}-\left(p^{2}-m^{2}\right)^{2}}.\label{eq: 0.09}
\end{equation}
We can proceed similarly for the gauge field, and find for the energy eigenvalues,
\begin{equation}
k_{0}\equiv\pm\omega_{k}=\pm\sqrt{M^{2}k^{2}+k^{4}},\label{eq: 0.10}
\end{equation}
and the solutions
\begin{equation}
A_{\mu}\left(x\right)=\sum_{i}\int\frac{d^{3}k}{\left(2\pi\right)^{\frac{3}{2}}}\left(\frac{1}{2\omega_{k}}\right)^{\frac{1}{2}}\left[a_{i}\left(k\right)\epsilon_{\mu}^{i}\left(k\right)e^{-ikx}+a_{i}^{\dagger}\left(k\right)\epsilon_{\mu}^{i}\left(-k\right)e^{ikx}\right]. \label{eq: 0.11}
\end{equation}
However, in order to define the photon propagator, we must impose a gauge condition, which reads a Lorentz-like condition
\[
\Omega\left[A\right]=\partial_{0}A_{0}-\left(-\Delta+M^{2}\right)\partial_{k}A_{k}=0,
\]
this leads to a non-local gauge-fixing term in the Lagrangian density \cite{ref22}, $\left(\xi=1\right)$,
\begin{equation}
\mathcal{L}_{gf}=-\left(\partial_{0}A_{0}-\left(-\Delta+M^{2}\right)\partial_{k}A_{k}\right)\frac{1}{2\left(-\Delta+M^{2}\right)}\left(\partial_{0}A_{0}-\left(-\Delta+M^{2}\right)\partial_{j}A_{j}\right),\label{eq: 0.12}
\end{equation}
but to a well-behaved expression for the propagator:
\begin{align}
i\mathcal{D}_{00}\left(k_{0},k\right) & =D_{00}\left(k_{0},k\right)=\frac{M^{2}+k^{2}}{k_{0}^{2}-M^{2}k^{2}-k^{4}},\label{eq: 0.13a}\\
i\mathcal{D}_{ij}\left(k_{0},k\right) & =D_{ij}\left(k_{0},k\right)=-\frac{\delta_{ij}}{k_{0}^{2}-M^{2}k^{2}-k^{4}},\label{eq: 0.13b}
\end{align}
with no off-diagonal components. Nevertheless, it was shown in the Ref.\cite{ref8} that off-diagonal components are generated after quantum corrections. We can read from the Lagrangian density \eqref{eq: 0.1} that the $z=2$ QED presents three vertices: \emph{two three-point} vertices,
\begin{equation}
\overline{\psi}A_{0}\psi\rightarrow ig\gamma_{0},\quad\overline{\psi}A_{k}\psi\rightarrow-ig\left(\gamma_{k}M+2p_{k}^{\left(\psi\right)}+p_{k}^{\left(A\right)}\right),
\end{equation}
which can be conveniently rewritten as in a compact form,
\begin{equation}
\left(\overline{\psi}A_{0}\psi,\overline{\psi}A_{k}\psi\right)\rightarrow\Lambda_{a}\left(q\right)=ig\left(\gamma_{0},-\gamma_{k}M-q_{k}\right),\label{eq: 0.14} 
\end{equation}
where we have introduced, by means of convenience in notation, the index $a=0,...,3$, which should not be confused with spacetime index, and $q_{k}\equiv2p_{k}^{\left(\psi\right)}+p_{k}^{\left(A\right)}$; and \emph{one four-point} vertex, $\overline{\psi}\psi A_{i}A_{j}\rightarrow -2ig\delta_{ij}$. But as we are here interested in the Bhabha scattering, we shall need to consider only \eqref{eq: 0.14}.

\section{Bhabha scattering}
\label{sec:2}

Collision experiments in high energy physics provide has been served as a ground framework for testing many theoretical proposals in QFT. Moreover, Bhabha scattering, $e^+ e^- \rightarrow e^+ e^-$, is one of the most fundamental reactions in QED processes \cite{ref11}, and its experimental data has been extensively used in order to obtain bounds in the context of Lorentz violating models \cite{ref14,ref7,ref8} as well as in the generalized electrodynamics \cite{ref9}.

\begin{figure*}[tbp]
\begin{center}
\includegraphics[scale=0.35]{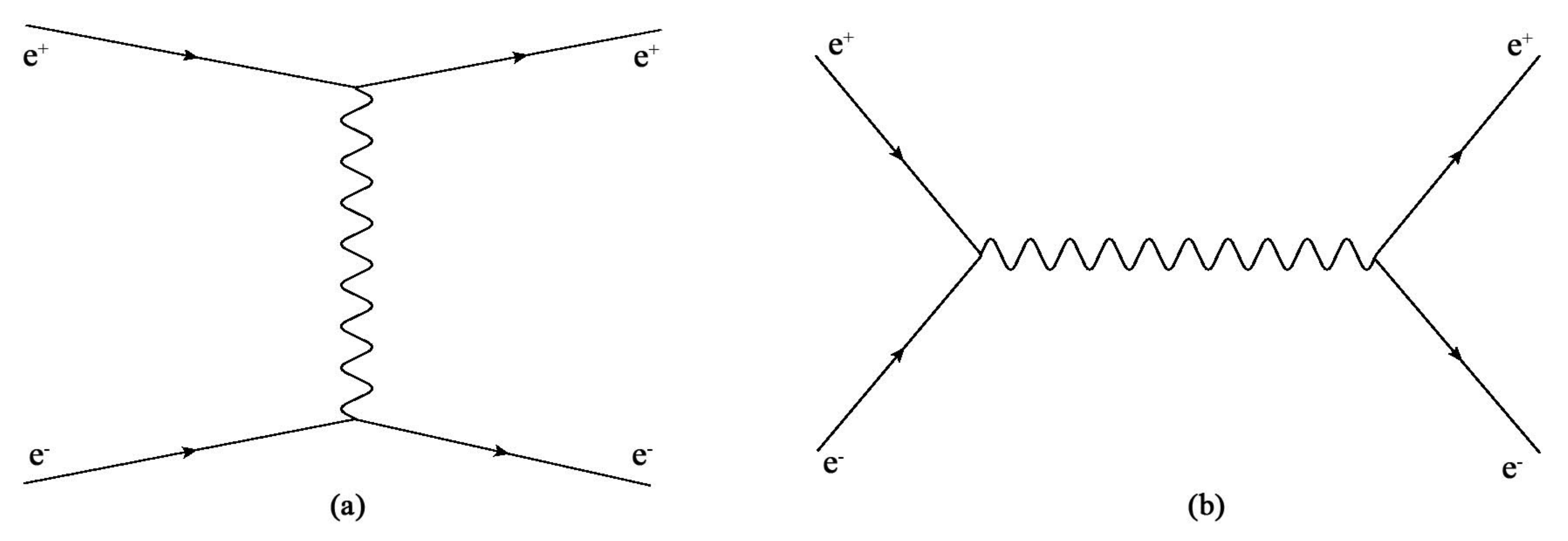}
\end{center}

\caption{Bhabha scattering: (a) photon exchange and (b) pair annihilation process.}
\label{fig1}
\end{figure*}

Let us consider the following description as for our system depicted at the Fig.\ref{fig1}: the spin and momenta for the ingoing $u_{r_{i}}\left(p_{1}\right)$ and outgoing $\overline{u}_{r_{f}}\left(p_{2}\right)$ electrons are $\left(r_{i},p_{1}\right)$ and $\left(r_{f},p_{2}\right)$, respectively, while the spin and momenta for the ingoing $\overline{v}_{s_{i}}\left(q_{1}\right)$ and outgoing $v_{s_{f}}\left(q_{2}\right)$ positrons are $\left(s_{i},q_{1}\right)$ and $\left(s_{f},q_{2}\right)$, respectively.  Nonetheless, it should be remarked that, because the photon propagator does not have off-diagonal components, we shall not have a mixed contribution between the temporal $\Lambda_{0}$ and spatial vertices $\Lambda_{k}$ in our calculation of the differential cross-section for the Bhabha scattering. From that we may now evaluate the total contribution for the transition amplitude, to lowest order,
\begin{equation}
i\mathcal{M}=i\mathcal{M}^{s}+i\mathcal{M}^{t},
\end{equation}
that corresponds to sum of the scattering diagram (s-channel)
\begin{equation}
i\mathcal{M}^{s}=-Ng^{2}\overline{v}_{s_{i}}\left(q_{1}\right)\Lambda_{a}\left(p_{1}-q_{1}\right)u_{r_{i}}\left(p_{1}\right)D_{ab}\left(p_{1}+q_{1}\right)\overline{u}_{r_{f}}\\
\left(p_{2}\right)\Lambda_{b}\left(p_{2}-q_{2}\right)v_{s_{f}}\left(q_{2}\right),\label{eq: 1.0}
\end{equation}
and of annihilation diagram (t-channel)
\begin{equation}
i\mathcal{M}^{t}=Ng^{2}\overline{v}_{s_{i}}\left(q_{1}\right)\Lambda_{a}\left(q_{1}\\
+q_{2}\right)v_{s_{f}}\left(q_{2}\right)D_{ab}\left(p_{1}-p_{2}\right)\overline{u}_{r_{f}}
\left(p_{2}\right)\Lambda_{b}\left(p_{1}+p_{2}\right)u_{r_{i}}\left(p_{1}\right),\label{eq: 1.1}
\end{equation}
where we have defined
\begin{equation}
N=\left(\frac{p_{1}^{2}-m^{2}}{E_{p_{1}}}\right)^{\frac{1}{2}}\left(\frac{p_{2}^{2}-m^{2}}{E_{p_{2}}}\right)^{\frac{1}{2}}\left(\frac{q_{1}^{2}-m^{2}}{E_{q_{1}}}\right)^{\frac{1}{2}}\left(\frac{q_{2}^{2}-m^{2}}{E_{q_{2}}}\right)^{\frac{1}{2}}.
\end{equation}
In order to evaluate the cross section for the Bhabha scattering, we shall now compute $\left|i\mathcal{M}\right|^{2}$ by taking an average over the spin of the incoming particles and sum over the outgoing particles
\begin{equation}
\left|i\mathcal{M}\right|^{2}\rightarrow\frac{1}{4}\sum_{r_{i},s_{i}}\sum_{r_{f},s_{f}}\left|i\mathcal{M}\right|^{2}.
\end{equation}
In this way, it follows
\begin{equation}
\frac{1}{4}\sum_{r_{i},s_{i}}\sum_{r_{f},s_{f}}\left|i\mathcal{M}\right|^{2}=\frac{1}{4}\sum_{spin}\left|\mathcal{M}^{s}\right|^{2}+\frac{1}{4}\sum_{spin}\left|\mathcal{M}^{t}\right|^{2}+\frac{1}{4}\sum_{spin}\left(\mathcal{M}^{s}\right)^{*}\mathcal{M}^{t}+\frac{1}{4}\sum_{spin}\left(\mathcal{M}^{t}\right)^{*}\mathcal{M}^{s}.\label{eq: 1.2}
\end{equation}
Although the calculation of each one of these terms is straightforward, the details are lengthy, but can be accomplished by using the completeness relations \eqref{eq: 0.6b} and by evaluating the traces of Dirac matrices products. Furthermore, as our main goal is to consider the behavior of the scattering process in the high energy limit, we shall take $m^2=0$ in our calculation. Nevertheless, in order to complete the calculation is interesting to express the kinematic variables in the center-of-mass frame, in which
\begin{equation}
p_{1}=\left(E,\mathbf{p}\right),\quad q_{1}=\left(E,-\mathbf{p}\right),\quad p_{2}=\left(E,\mathbf{q}\right),\quad q_{2}=\left(E,-\mathbf{q}\right).
\end{equation}
Besides, due energy conservation: $E_{i}=E_{f}\rightarrow\left|\mathbf{p}\right|=\left|\mathbf{q}\right|$, and also we define $\theta$ as the center-of-mass scattering angle, i.e. $\left(\textbf{p}.\textbf{q}\right)=p^2 \cos \theta$. Hence, the resulting expression for all these contributions read, for the s-channel contribution,
\begin{gather}
\frac{1}{4}\sum_{spin}\left|\mathcal{M}^{s}\right|^{2}  =\frac{4g^{4}}{\left(4\varphi\right)^{4}M^{4}}I_{s}\left(E,M,\theta\right),\\
\frac{1}{4}\sum_{spin}\left|\mathcal{M}^{t}\right|^{2}  =\frac{g^{4}}{8M^{4}}\frac{1}{\sin^{4}\left(\frac{\theta}{2}\right)\left[1+2\left(1-\Delta\right)\sin^{2}\left(\frac{\theta}{2}\right)\right]^{2}}\frac{I_{t}\left(E,M,\theta\right)}{\left(1-\Delta\right)^{2}\left(2\varphi\right)^{2}},\\
\frac{1}{4}\sum_{spin}\left(\mathcal{M}^{s}\right)^{*}\mathcal{M}^{t}+\frac{1}{4}\sum_{spin}\left(\mathcal{M}^{t}\right)^{*}\mathcal{M}^{s}  =\frac{g^{4}}{8M^{4}}\frac{1}{\sin^{2}\frac{\theta}{2}\left(1+2\left(1-\Delta\right)\sin^{2}\frac{\theta}{2}\right)}\frac{I_{st}\left(E,M,\theta\right)}{\left(1-\Delta\right)\left(4\varphi\right)^{3}},
\end{gather}
where we have defined the quantities $I_{s} $, $I_{t}$ and $I_{st}$, in the following way
\begin{align}
I_{s}  \left(E,M,\theta\right)=&23+104\varphi+76\varphi^{2}-23\Delta-58\varphi\Delta+4\varphi\cos\theta\left(1-\varphi\right)\notag\\
&+\cos2\theta\left(23+84\varphi+32\varphi^{2}-23\Delta-38\varphi\Delta\right),
\end{align}
while
\begin{align}
I_{t}\left(E,M,\theta\right)  =&2\left(63\left[1-\Delta\right]+\varphi\left(296-170\Delta
+\varphi\left[299+32\varphi-72\Delta\right]\right)\right)\nonumber \\
 & +\cos\theta\left(135\left[1-\Delta\right]+\varphi\left(621-351\Delta+\varphi
 \left[536-96\Delta\right]\right)\right)\notag \\
 &+\cos2\theta\left(61\left[1-\Delta\right]+2\varphi\left(154-93\Delta+8\varphi
 \left[23+4\varphi-7\Delta\right]\right)\right)\nonumber \\
 & +\cos3\theta\left(5\left(1-\Delta\right)+\varphi\left(19+8\varphi-9\Delta\right)\right)
 +\cos4\theta\left(\left(1-\Delta\right)+2\varphi\left(2+\varphi-\Delta\right)\right).
\end{align}
and 
\begin{align}
I_{st}\left(E,M,\theta\right) =&142\left(1-\Delta\right)+\varphi\left[747-463\Delta+2\varphi\left(385+72\varphi-16\Delta\right)\right]\nonumber \\
 & +2\cos\theta\left[154\left(1-\Delta\right)+\varphi\left(1241-933\Delta+2\varphi
 \left[1063+72\varphi-252\Delta\right]\right)\right]\nonumber \\
 & +\cos2\theta\left[182\left(1-\Delta\right)+\varphi\left(783-419\Delta+2\varphi\left[349+72\varphi
 -100\Delta\right]\right)\right]\nonumber \\
 & +8\cos3\theta\left[11\left(1-\Delta\right)+\varphi\left(52-30\Delta+47\varphi-9\varphi\Delta\right)\right]\nonumber \\
 &+4\cos4\theta\left[3\left(1-\Delta\right)+2\varphi\left(7-4\Delta+\varphi\left[6-\Delta\right]\right)\right].
\end{align}
with $\varphi\equiv\frac{E^2}{M^4}$ and $\Delta\equiv\sqrt{1+4\varphi}$. Nevertheless, we have that the differential cross section (in natural units) is given by
\begin{equation}
\frac{d\sigma}{d\Omega}=\frac{E^{2}}{\pi^{2}}\frac{1}{4}\sum_{r_{i},s_{i}}\sum_{r_{f},s_{f}}\left|i\mathcal{M}\right|^{2},
\end{equation}
then, with the previous results one obtain.
\begin{align}
\frac{d\sigma}{d\Omega} =&\frac{\alpha^{2}}{4\varphi^{3}}I_{s}\left(E,M,\theta\right)+\frac{\alpha^{2}}{2\left(1-\Delta\right)^{2}\varphi}\frac{I_{t}\left(E,M,\theta\right)}{\sin^{4}\left(\frac{\theta}{2}\right)\left[1+2\left(1-\Delta\right)\sin^{2}\left(\frac{\theta}{2}\right)\right]^{2}}\nonumber \\
 & +\frac{\alpha^{2}}{32\left(1-\Delta\right)\varphi^{2}}\frac{I_{st}\left(E,M,\theta\right)}{\sin^{2}\frac{\theta}{2}\left(1+2\left(1-\Delta\right)\sin^{2}\frac{\theta}{2}\right)},\label{eq: 1.15}
\end{align}
where $\alpha=\frac{g^{2}}{4\pi}$. To conclude this section, based on experimental grounds, we may determine limit bounds for the parameter $M$ in the case previously evaluated. For Bhabha scattering, the experimental data on precision tests in QED are readily available in Ref.\cite{ref10}, and small deviations on the QED tree results may be expressed in the form
\begin{equation}
\delta\equiv\left(\frac{d\sigma}{d\Omega}\right)\left/\left(\frac{d\sigma}{d\Omega}\right)_{QED}\right.-1\approx\pm\frac{3E_{cm}^{2}}{\Lambda_{\pm}^{2}},
\end{equation}
where $E_{cm}=2E=29~GeV$ and $\Lambda_{+}=200~GeV$ with $95\%$ confidence level; actually, $\Lambda$ and is a parameter representing possible experimental departures from the theoretical predictions. Considering the leading corrections from \eqref{eq: 1.15} with $\left|\cos\theta\right|<0.55$ \cite{ref10}, we can show that the magnitude of these corrections are of order $\varphi=\frac{E^{2}}{M^{4}}$, and therefore when we compare them in $\delta$ we obtain the bound limit $M^{2}\geq10^{10}eV$, for $\Lambda=200~GeV$.

\section{Concluding remarks}

\label{sec:3}

In this paper we have discussed the Bhabha scattering in the framework of Lifshitz field theory, the $z=2$ quantum electrodynamics. This theory has shown to be rather interesting both in classical and quantum context, and also phenomenological viable Lifshitz extension of QED. Although it is a super-renormalizable field theory, it provided rich outcomes in the radiative corrections.

Recently, due to the accuracy of the experimental data of Bhabha scattering, interesting studies have used this data in a stringent way in setting bound values on theory's free parameters have appeared in the literature, such as in the case of Lorentz violating theories and generalized electrodynamics. Nonetheless, we have started by discussing the classical theory in order to construct the free solution for the fermionic fields, since they now obey a modified Dirac equation and possess a different dispersion relation. These results were necessary in order to compute correctly the S-matrix elements (transition amplitude). In particular, we were interested in the high-energy behavior of the cross section, hence we took $m^2=0$ in our calculation. Afterwards, we have compared the leading corrections from the $z=2$ QED with small deviations from the experimental data of the Bhabha scattering $95\%$ confidence level, finding that for values $M^{2}\geq10^{10}eV$ the theory is compatible with the experimental data.

Due to the richness of outcomes that $z=2$ QED has provided, we believe that may be there are many other issues that should be discussed. For instance, we may cite many investigations on Lifshitz-like theories concerning in the study of the effective potential considering an arbitrary value of the critical exponent $z$ and for different kind of couplings and fields \cite{ref15a,ref15b,ref15c,ref15d,ref15e,ref15f,ref20}. In particular, an extension of our studies for $z=2$ QED for the finite-temperature case \cite{ref21} certainly would be very important, since the high-temperature corrections provide a description of a large class of interesting phenomena and these contributions may be put in a closed form expression \cite{ref27a,ref27b}. These issues and others will be further elaborated, investigated and reported elsewhere.

\subsection*{Acknowledgments}

R.B. thanks FAPESP for full support.


\end{document}